# Simulations of $Nb_3Sn$ Layer RF Field Limits Due to Thermal Impedance

Paulina Kulyavtsev, Grigory Eremeev, Sam Posen

**Abstract**
$Nb_3Sn$ performance in RF fields is limited to fields far below its superheating critical field. Poor thermal conductivity of $Nb_3Sn$ has been speculated to be the reason behind this limit. In order to better understand the contribution of $Nb_3Sn$ thermal conductivity to its RF performance, we simulated numerically with Matlab program*(based on SRIMP and HEAT codes) the limiting fields under different realistic conditions. Our simulations indicate that limiting fields observed presently in the experiments with RF fields cannot be explained by the thermal impedance of $Nb_3Sn$ alone. The results change significantly in the presence of higher losses due to extrinsic mechanisms.

**Background:**
$Nb_3Sn$ is a material of interest due to potential in promising performance and cost reduction. It is implemented as a coating of a thin film inside existing Nb cavities, followed by an annealing in vacuum [1]. The goal is to operate at higher temperatures and higher surface magnetic fields.

Operating temperatures of interest are 2K and 4K. 2K is the current typical temperature for operation of Nb cavities in accelerators, and 4K is a potential operating condition for $Nb_3Sn$ which would significantly reduce operating costs due to cooling [1].

Although superconductors are capable of conducting high DC currents without any losses, RF currents in SRF cavities dissipate some heat, typically on the scale of watts, which has to be removed by liquid helium. Heat must go through the wall of the cavity to be dissipated by the helium bath.

The goal of this project was to simulate the process of heat flow from the vacuum side to the He side, and show how it is affected by the thermal impedance of the $Nb_3Sn$ layer. By varying the thermal impedance, e.g., via the thickness of the $Nb_3Sn$ layer, we estimate the RF field at which thermal runaway occurs under different circumstances. For this project, incremental steps are taken to use the existing HEAT model of Nb and modify it to represent $Nb_3Sn$, and then again for a $Nb_3Sn$-Nb bilayer.

$Nb_3Sn$ is known to have significantly poorer thermal conductivity than Nb [2]. Our simulations were designed to improve understanding of the field limit in $Nb_3Sn$ cavities to see whether it may be due to this poor thermal conductivity causing thermal instability. Primarily, two cases could be considered in this simulation; one dimensional or Global Thermal Instability (GTI), and two dimensional overheating which can be used to study the effect of defects. In this paper, we focus on GTI, assuming a relatively defect-free surface. While thermal instability is reasonably well understood for niobium, it is not as well addressed for $Nb_3Sn$.

A critical aspect of cavity performance is the ability to remove the heat generated on the inside surface of cavity by cooling the other side. Many factors contribute to the ability to remove this heat: the field on the surface, the surface resistance, the conduction of the heat through the cavity wall, Kapitza conduction at the outer wall-bath interface, as well as the helium bath temperature and helium convection. Thermal instability could result from the collective contribution of these factors in hindering the removal of heat from the RF surface. This is important for the simulation as the impact of these contributions may be indirectly compared to how extreme values of each factor impacts the projected performance and to elucidate which factors in which conditions dominate the contribution to thermal instability. Effects of factors such as Kapitza conductance and thickness of $Nb_3Sn$ layer were evaluated through the simulation. In our simulation, the one-dimensional analysis was done by assuming the $Nb_3Sn$ layer is an infinite plane in a lateral direction, limiting heat transfer to only the z direction.

*Figure 1: Increasing thicknesses of $Nb_3Sn$ in a specific field. At some point, there is a failing thickness; global thermal runaway occurs as heat is unable to be removed. This is the thermal runaway critical thickness.*

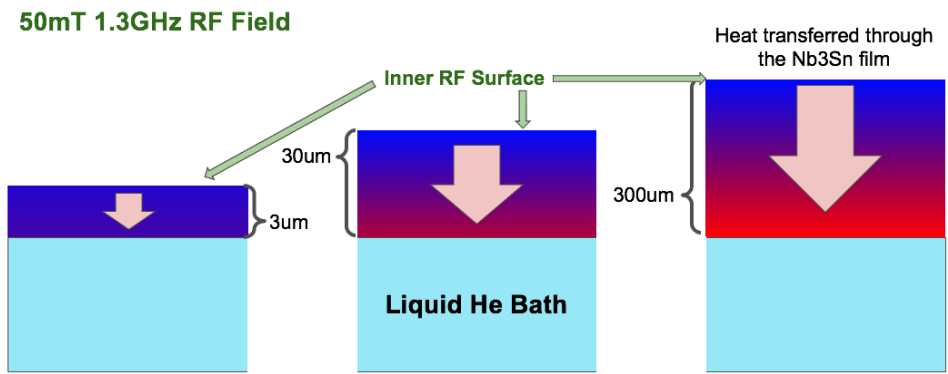

Our simulation is of an ideal situation, but defects can further complicate the story. The contribution of defects would be simulated by a two-dimensional model looking for the defect-induced thermal instability. A defect, which is more lossy than the surrounding material, becomes an area of excess heating, and this additional heat contribution greatly increases the power dissipation driving thermal runaway; therefore, defects are of interest to simulate to better understand their impact on thermal conductivity and the successful removal of heat from the material.

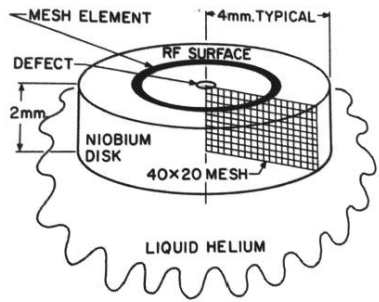

*Figure 2: The model used by the HEAT program to simulate defects using a two-dimensional mesh, relying on the symmetry of the problem to simulate the onset of thermal breakdown [3].*

**Simulation Method**

The following simulations built on the work of Xie and Meng [4]. First, the program was used to evaluate Nb case. The results were found consistent with published results.

*Exploration of a $Nb_3Sn$ Case*
With the program having been benchmarked against the niobium case, separate cases were added to the existing Niobium-based functions to model $Nb_3Sn$. Specifically, $Nb_3Sn$ surface resistance and thermal conductivity were added to reflect the new material under study.

For $Nb_3Sn$ thermal conductivity, a polynomial fit was derived from published data [2]. The fit as well as the data from [2] are shown in Figure 3. Using the polynomial fit instead of inline approximation has helped to considerably speed up the calculation time.

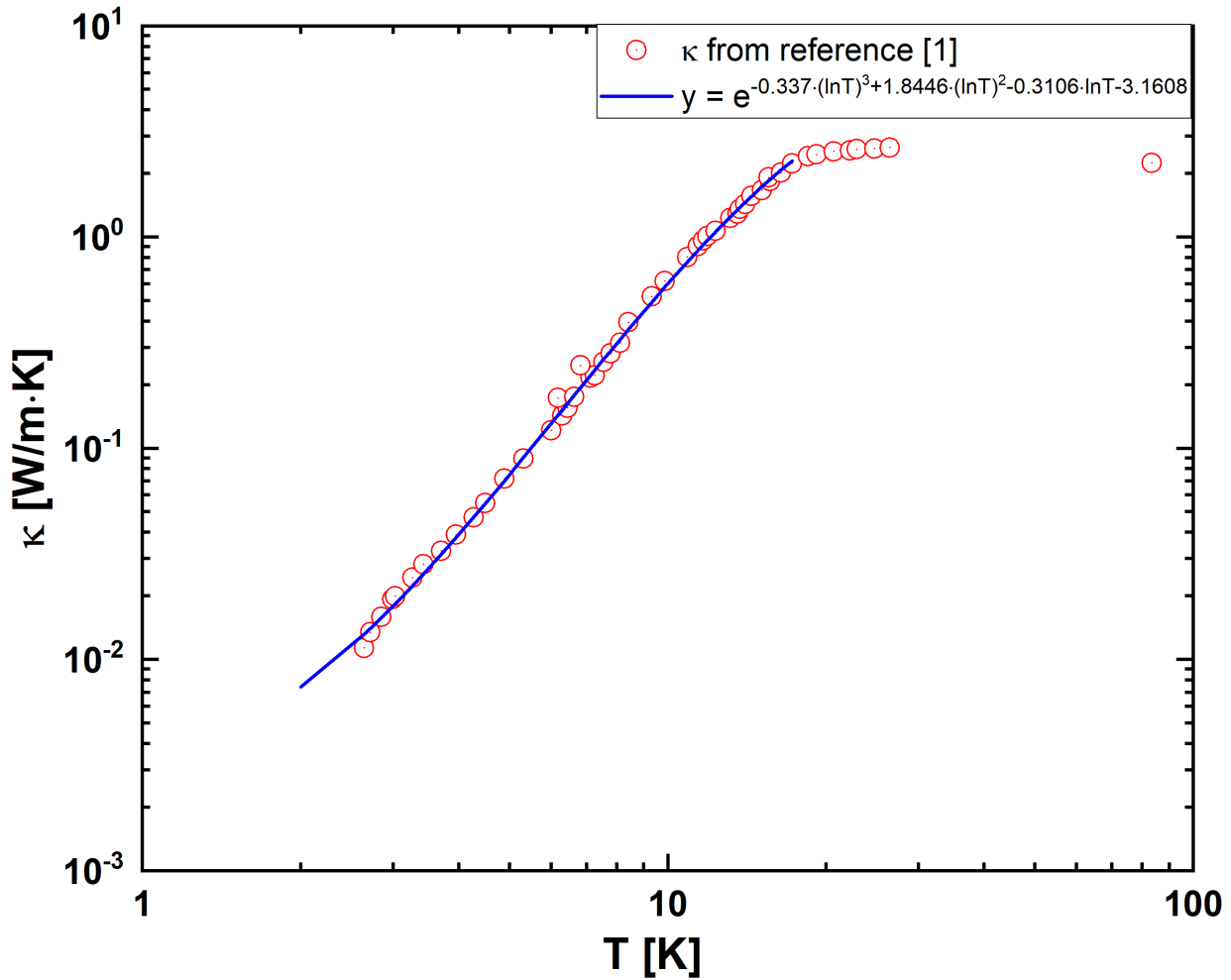

Figure 3: Thermal Conductivity vs. Temperature for Nb3Sn material from [2]. Polynomial fits to the thermal conductivity from [2] were used in the program.

For BCS resistance, SRIMP was first used to calculate RF surface resistance of Nb3Sn in the relevant temperature range [5]. SRIMP-calculated $Nb_3Sn$ BCS resistance was fit with an explicit function that was then used in the numerical simulations for surface resistance calculations to significantly reduce the calculation time, Fig. 4.

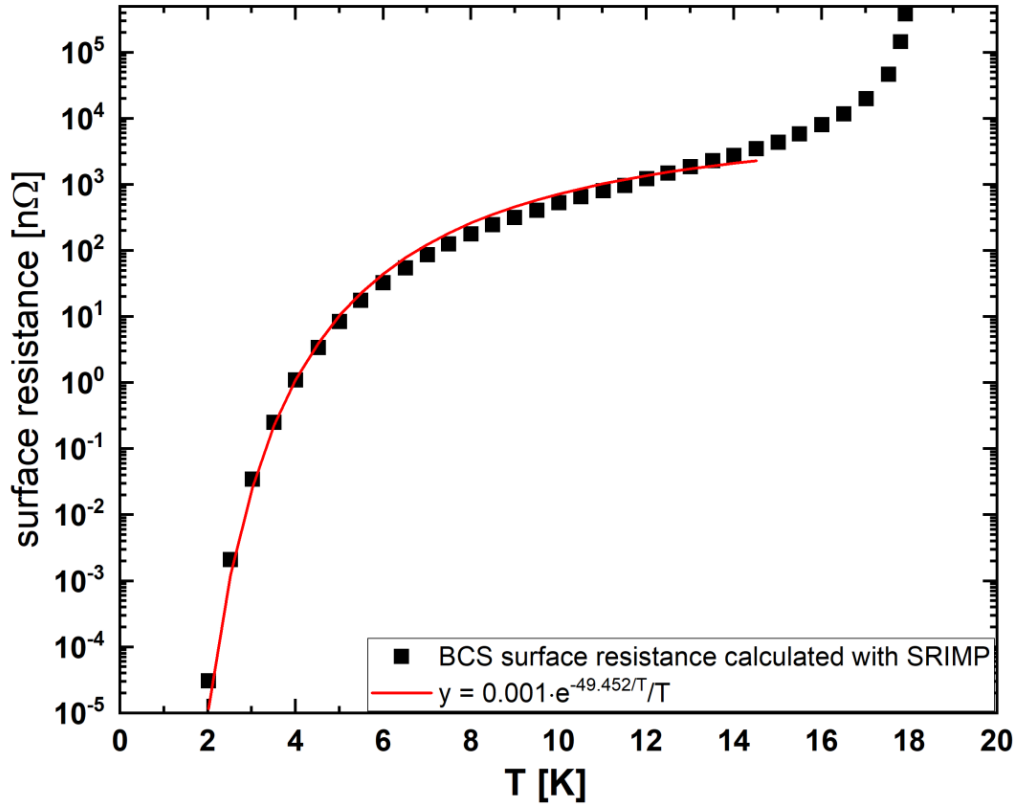

*Figure 4: BCS resistance case was written based on the SRIMP calculations, using polynomial fit to the calculated resistance.*

In Figure 4 we show both $Nb_3Sn$ surface resistance, calculated using SRIMP, and the fitting function. The fitting function is:

$$R_s(T) = A\frac{e^{-\frac{B}{T}}}{T}$$

, where the fitting parameters are A = 1.0·10$^{-3}$ Ω·K, B = 49.452 K.

Calculations that we report here were setup in the following way. For each parameter set we simulated the temperature rise as a function of material thickness. For each applied field, we eventually arrived at the thickness above which no stable solution could be found in our simulations. The maximum thickness at which solution still existed was called the "critical" thickness for the given parameter set. We then find the "critical" thickness for varying parameter sets.

**Results**

*Kapitza conductance*

One of our first assumptions was that Kapitza conductance is not the main factor limiting $Nb_3Sn$ performance, and so we simplified our simulation by excluding it from our simulation. Before removing thermal impedance due to Kapitza boundary resistance from our simulation, we first evaluated Kaptiza contribution to field limits by simulating critical film thickness versus field strength for several different Kapitza resistances. Several different Nb cases were used for this simulation: annealed Niobium, unannealed niobium, and annealed niobium*1000, which means that the Kapitza resistance for annealed niobium was further increased by a factor of 1000.

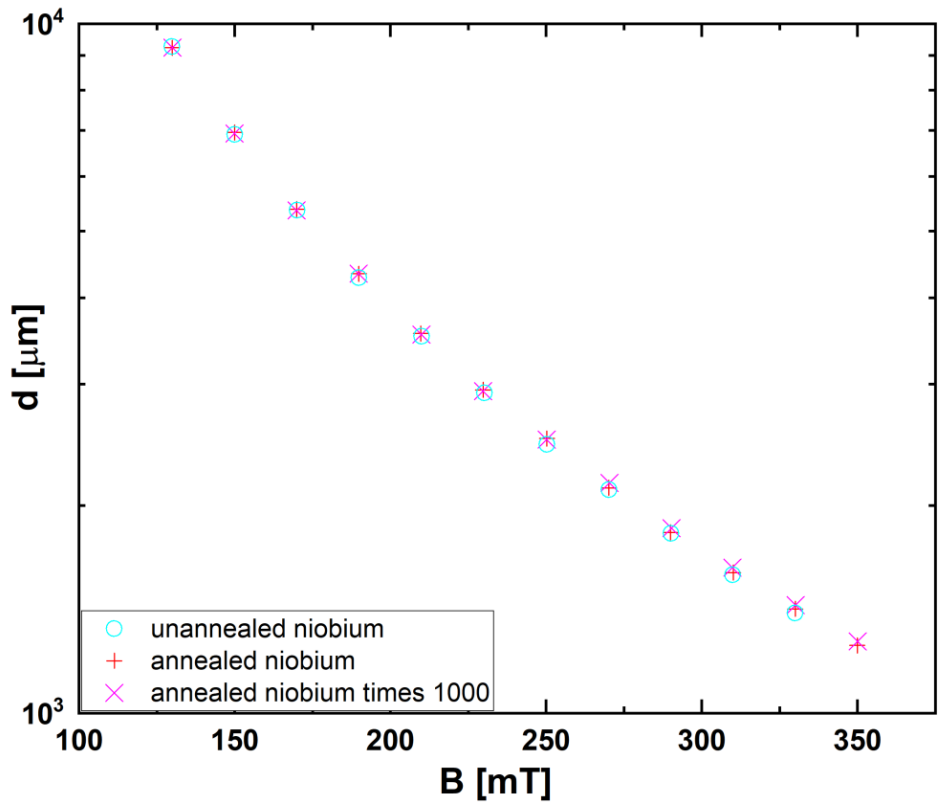

*Figure 5: Critical thickness as a function of field for different Kapitza resistances. Simulations were done for 2K with residual resistance of 10 nOhm.*

The results of Kapitza resistance variation is that it has negligible effect. High Kapitza conductance means a very low Kapitza resistance and, if Kapitza boundary played a role in the overall heat transport, we would expect different results for different Kapitza resistances. As shown in Fig.5., varying Kapitza conductance did not affect the results, even when it was increased by three orders or magnitude. This shows that removal of heat from the RF surface is not significantly impeded by the cavity surface - liquid He interface for a typical range of values.

*Thermal Conductivity*

To check the impact of thermal conductivity, we simulated the critical thickness for different thermal conductivities at two different helium bath temperatures, 2 K and 4 K. As a baseline thermal conductivity, we used the thermal conductivity from [2], also shown in Fig 3. The critical thickness simulations were then done with thermal conductivity reduced by a factor of 2, increase by a factor of two, and a factor of three. The results are shown in Fig 6.

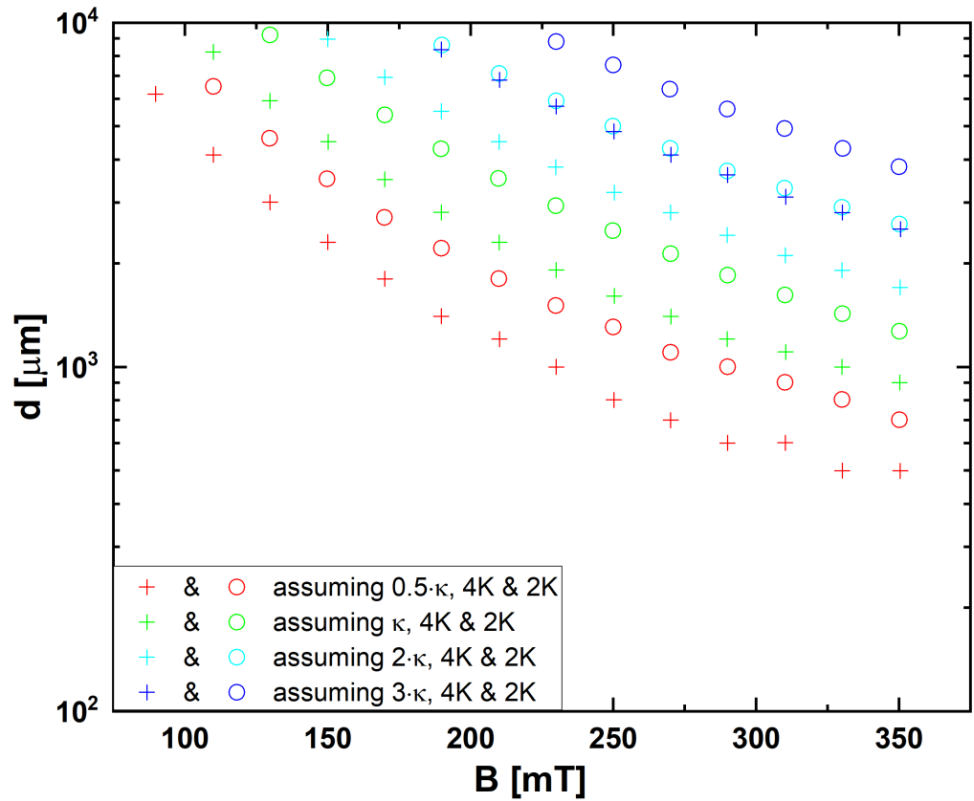

*Figure 6: Critical thickness simulation results for varying thermal conductivities and two different bath temperatures. Kapitza conductance for annealed niobium increased by a factor of 1000 and the residual surface resistance of 10 nOhm were used.*

As expected, thermal conductivity has significant impact on the thermal impedance. The critical thickness scales approximately linearly with the thermal conductivity at both simulated helium bath temperatures.

To check how the helium bath temperature affects the critical thickness, we simulated the critical thickness as a function of applied field for three different helium bath temperatures, 2 K, 4 K, and 6 K. As was discussed earlier, thermal simulation for varying Kapitza resistances indicated that Kapitza resistance contribution to overall thermal impendence is negligible, so the same Kapitza resistance of unannealed niobium was used for different outer wall temperatures. The results of this simulation are shown in Fig 7.

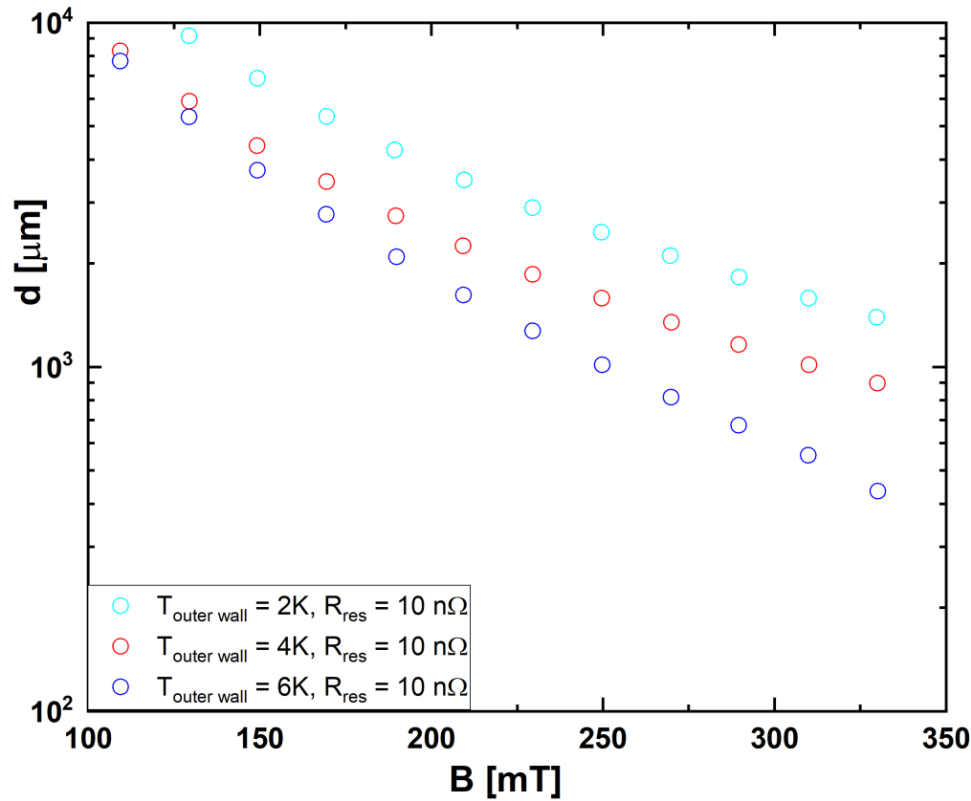

*Figure 7: Critical thickness simulation results for different helium bath temperatures. Kapitza conductance of unannealed Niobium and the residual surface resistance of 10 nOhm were used.*

The lower outer wall temperature improves the stability of the material. And the effect of the outer wall temperature increases with the applied field. Decreasing the outer wall temperature from 6 K to 2 K improves the critical thickness by almost a factor of three at higher fields.

Finally, we simulated the effect of the residual resistance on the critical thickness, Fig 8. For this simulation four different surface resistances, 1 nOhm, 10 nOhm, 100 nOhm, and 1000 nOhm, were used. For this simulation Kapitza resistance of annealed niobium increased by a factor of 1000 was used.

For the highest simulated surface resistance of 1000 nOhm, the critical thickness drops to about 60 µm at the highest fields, which is comparable to the thickest $Nb_3Sn$ grown on SRF cavities. At low fields of about 100 mT, the critical thickness is still at or above 1 mm.

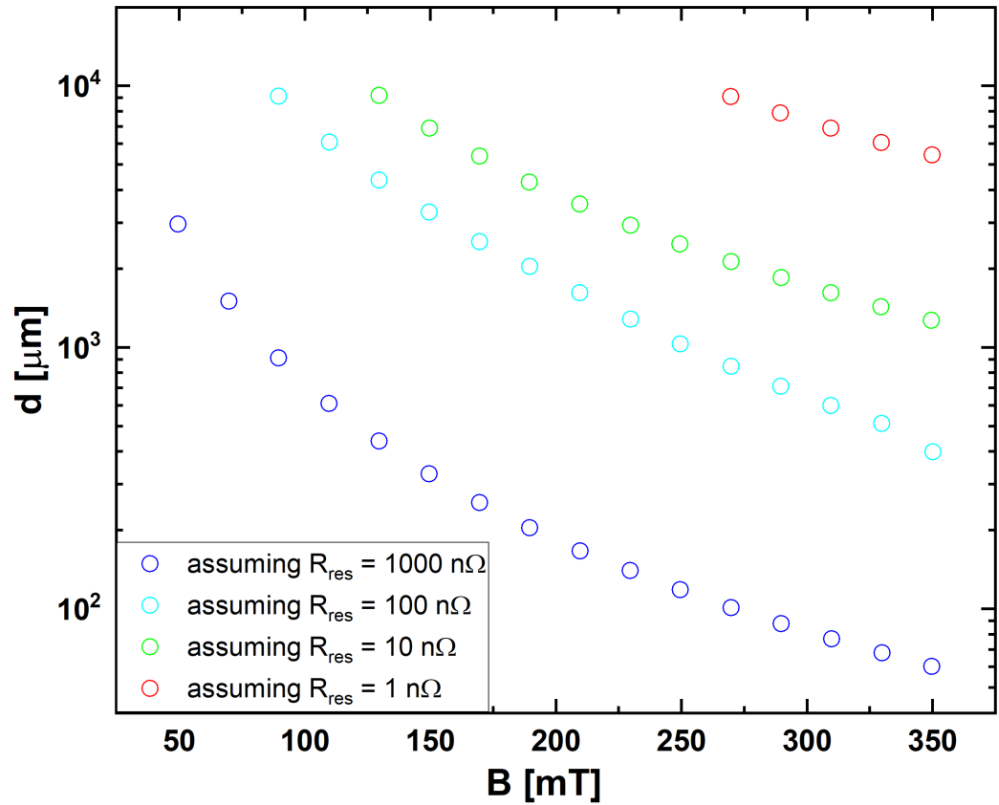

*Figure 8: Critical thickness simulation results for different residual surface resistance. Residual resistance has significant impact on the thermal breakdown.*

**Discussion**

With the goal to explore how different material properties affect the maximum material thickness, which still allows for the RF surface to be adequately cooled by liquid helium at the outer wall at a given field, we first check the influence of Kapitza resistance. Under the assumption of the thermal resistance value reported in [2] and the residual surface resistance of 10 nOhm, we simulated the critical thickness vs field for three different values of the Kapitza resistance: annealed, unannealed, and also Kapitza resistance of annealed niobium reduced by a factor of 1000, which represented the case of extremely low Kapitza resistance. The results of the simulation shown in Fig. 5. demonstrated that there is little difference in the critical thickness for all simulated fields between the Kapitza resistance reduced by a factor of 1000 and the other two cases. The simulation results show that Kapitza resistance has little impact on the overall thermal impedance and can be ignored in the simulations.

These simulations also showed that the critical thickness for which thermal runaway occurs is, at least, several millimeter even for a the fields as high as 200 mT, which is significantly higher than the highest fields reached in the best $Nb_3Sn$-coated SRF cavities[6,7]. The typical thickness of Nb$_3$Sn coating on RF surface cavities is 1-3 µm, which in the light of present results significantly thinner than the thickness, which will impact heat transfer to the helium bath.

As we expected the critical thickness scales approximately linearly with thermal conductivity, Fig. 6. For a typical thermal conductivity value [2] the critical thickness far exceeds the typical coating thickness in modern SRF cavities, which is consistent with simulation reported earlier [8]. The thermal conductivity must be suppressed by a factor of 10000 to become the key limiting factor for applied RF fields below 100 mT.

The other factor that has significant impact to the critical thickness is the surface resistance, Fig. 8. The critical thickness drops by about two orders of magnitude when the residual surface resistance is increased from nOhm to 1000 nOhm. Three orders of magnitude may seem as a significant change in the surface resistance, but, in fact, there are many well-known physical mechanisms that can cause such degradation. For example, a normal-conducting inclusion will have a surface resistance of several mOhm. Such small localized defects seem to be one of the likely candidates behind present limitations.

These simulations, where we varied potential contributors to GTI, suggest there is an external limitation, not intrinsic to the material. Perhaps defects are the limitation; the next avenue of exploration would be to look into thermal runaway due to defects varying factors such as defect resistance and defect size. A further advancement to be made on these simulations would be to simulate two materials, Nb$_3$Sn film on Nb, and again evaluate where the thermal runaway occurs.

**Conclusion**

We numerically simulated different thicknesses of Nb$_3$Sn films in varying realistic conditions using a temperature dependent thermal conductivity. Our simulations indicate that limiting fields observed presently in the experiments with RF fields cannot be explained by the thermal impedance of Nb$_3$Sn alone. We found that the primary contribution to the thermal resistance was the Nb$_3$Sn thermal conductivity. The results change significantly in the presence of higher losses due to extrinsic mechanisms. At the magnitude of microns thickness, GTI is not a concern for state-of-the-art Nb$_3$Sn films. From this, there is interest in further simulation of defects as a two-dimensional case and Nb-Nb$_3$Sn bilayers to gain a better understanding of what that external limitation may be.

**Acknowledgements**